\documentclass[aps,10pt,twocolumn,prl,floats,floatfix]{revtex4-1}
\usepackage{verbatim}
\usepackage{listings}
\usepackage{graphicx}
\usepackage{color}
\usepackage{amsmath}
\usepackage{amssymb}
\usepackage[dvips]{epsfig}
\usepackage[T1]{fontenc}
\usepackage{shadow}
\usepackage{varioref}
\usepackage{epsfig}
\usepackage{array}
\usepackage{rotating}

\newcommand{\xRM}{\mathrm{RRMx}}
\newcommand{\RM}{\mathrm{RM}}
\newcommand{\RMin}{\mathrm{RRM}}

\newcommand{\aRMinavg}{\left<|\mathrm{RRM}|\right>}
\newcommand{\RMobs}{\mathrm{RM}_{\mathrm{obs}}}
\newcommand{\RMgal}{\mathrm{RM}_{\mathrm{gal}}}
\newcommand{\Lum}{L_{1.4\text{GHz}}}
\newcommand{\Lya}{Ly$\alpha$}

\newcommand{\be}{\begin{equation}}
\newcommand{\ee}{\end{equation}}
\newcommand{\bea}{\begin{eqnarray}}
\newcommand{\eea}{\end{eqnarray}}

\newcommand{\dd}{\text{d}}

\newcommand{\lp}{({\it lp})~}

\begin{document}
\title{New limits on extragalactic magnetic fields from rotation measures}

\author{M. S. Pshirkov$^{1,2,3}$, P. G. Tinyakov$^4$, F. R. Urban$^4$\\
$^{1}$ {\footnotesize{\it Sternberg Astronomical Institute, Lomonosov Moscow State University, Universitetsky prospekt 13, 119992, Moscow, Russia}}\\
$^{2}$ {\footnotesize{\it Institute for Nuclear Research of the Russian Academy of Sciences, 117312, Moscow, Russia}}\\
$^{3}$ {\footnotesize{\it Pushchino Radio Astronomy Observatory, 142290 Pushchino, Russia}}\\
$^{4}$ {\footnotesize{\it Universit\'e Libre de Bruxelles, Service de Physique Th\'eorique, CP225, 1050, Brussels, Belgium}}\\
}
\noaffiliation

\begin{abstract} 
We take advantage of the wealth of rotation measures data contained in the NRAO VLA Sky Survey catalogue to derive new, statistically robust, upper limits on the strength of extragalactic magnetic fields.  We simulate the extragalactic magnetic field contribution to the rotation measures for a given field strength and correlation length, by assuming that the electron density follows the distribution of Lyman-$\alpha$ clouds.  Based on the observation that rotation measures from distant radio sources do not exhibit any trend with redshift, while the extragalactic contribution instead grows with distance, we constrain fields with Jeans' length coherence length to be below 1.7~nG at the $2\sigma$ level, and fields coherent across the entire observable Universe below 0.65~nG.  These limits do not depend on the particular origin of these cosmological fields.
\end{abstract}

\maketitle

%%%%%%%%%%%%%%%%%%%%%%%%%%%%%%%%%%%%%%%%%%%%%%%%%%%%%%%
\section{\label{sec:mot}Motivation}
%%%%%%%%%%%%%%%%%%%%%%%%%%%%%%%%%%%%%%%%%%%%%%%%%%%%%%%

Is the Universe permeated by an all-encompassing magnetic field (MF)?   MFs are already observed in nearly all types of structures, from planets to galaxies to clusters of galaxies~\cite{2004NewAR..48..763V,Kronberg:1993vk,Durrer:2013pga,Widrow:2011hs}, but cosmological magnetic fields still remain elusive.  Nonetheless, understanding their characteristics has cardinal relevance in many fields in astroparticle physics and cosmology: propagation of ultra-high energy cosmic rays (UHECRs), structure formation, early and very early Universe models, physics beyond the Standard Model, radio-astronomy, and so on, see~\cite{2004NewAR..48..763V,Kronberg:1993vk,Durrer:2013pga,Widrow:2011hs,Kandus:2010nw,Subramanian:2015lua}.

One way cosmological MFs manifest themselves is by rotating the plane of polarisation of electromagnetic waves propagating from far away sources to the Earth.  The main idea of this work stems from a very simple observation we made in our previous paper~\cite{Pshirkov:2014lqa}: Faraday rotation measures (RMs) of distant objects do not show any evolution with redshift.  However, if there is an all-pervading, extragalactic MF (egMF) we do expect a quite pronounced change in the distribution of RMs with distance.  We can thus limit the strength of such a field by comparing simulated and observed RMs distributions.

Currently, the strongest upper limits on the strength, $\hat{B}$, of present-day egMFs come from microwave background observations~\cite{Planck2015_PMF} (see also~\cite{Barrow:1997mj,Trivedi:2013wqa}) and read $\hat{B} \lesssim 2.8$~nG for a coherence length $l_c = 1$~Mpc.  In the special case of a scale-invariant MF spectrum, these limits could be further lowered to 0.9~nG, if one accounts for their impact on the ionisation history of the Universe.  Notice that these limits apply only to primordial MFs, i.e., fields generated in the very early Universe, while, for instance, cosmological MFs could be generated at later stages by various astrophysical mechanisms (e.g.~\cite{Kronberg2001,Miniati2011,Vazza:2014jga}).

The limits coming from the analysis of RM data are less restrictive: according to~\cite{Blasi:1999hu} (see also~\cite{Kronberg:2007dy}), fields correlated on Mpc scales are bound to have $\hat{B} \lesssim 6$~nG.  A comprehensive overview of present and earlier constraints coming from RM observations can be found in~\cite{Durrer:2013pga}.

We devise a method for extracting the unknown errors from the data itself, and thanks to this new method and the new available RM data, we can improve on these limits by five times, and include a full treatment of their statistical significance: we find that egMFs with coherence lengths of about 1~Mpc and strengths above 1.7~nG (1.2~nG using only low-luminosity ({\it lp}) sources) are incompatible with current RM observations at the $2\sigma$ level; this limit becomes 0.65~nG (0.50~nG) if the egMF is coherent across the entire Universe.

%%%%%%%%%%%%%%%%%%%%%%%%%%%%%%%%%%%%%%%%%%%%%%%%%%%%%%%
\section{Method}\label{sec:met}
%%%%%%%%%%%%%%%%%%%%%%%%%%%%%%%%%%%%%%%%%%%%%%%%%%%%%%%

\paragraph*{Observations.} The plane of polarisation of a linearly polarised electromagnetic wave which moves through a magnetised plasma rotates by an angle $\varphi$ proportional to the square of the wavelength $\lambda$: $\varphi=\RM\,\lambda^2$, where
\be
\RM = 812\int_{D}^{0} \frac{n_\text{e}(z) B_{||}(z)}{(1+z)^2} \left|\frac{\dd l(z)}{\dd z}\right|\dd z \, . \label{RM}
\ee
Here $n_\text{e}$ is the density of free electrons measured in cm$^{-3}$, $B_{||}$ is the component of the MF (in $\mu$G) parallel to the line of sight $l(z)$, and $D$ is the distance to the source in kpc; here and everywhere $\RM$ is measured in $\text{rad}/\text{m}^2$.  Notice that in order to gain some knowledge about the MF, some independent estimates of $n_\text{e}$ are required.

The largest set of RM of extragalactic sources to date was compiled in~\cite{Taylor:2009} from the NRAO VLA Sky Survey (NVSS) data~\cite{Condon:1998iy}.  The total number of observed sources was 37,543, of which 4002 have known redshifts~\cite{Hammond:2012pn}.  From this set, we accepted only sources with galactic latitude $|b|>20^{\circ}$.

Generically, we can split the observed RM as $\RMobs=\RMgal+\RMin$, where the first term is the contribution of the regular MF of the Milky Way, and the second term stands for ``residual RM'', and encodes all other sources of RM once the local (regular) MF is subtracted: RM instrinsic to the source, measurement errors, turbulent galactic MF, and egMF ($\xRM$), see~\cite{Schnitzeler:2010ax} for an estimate of these components.

In order to disentangle redshift-dependent effects from those pertaining to the sources themselves, we computed their luminosity using, where available, the most recent spectral indices from~\cite{Farnes:2014a}; since not all sources have measured spectral indices, this reduced the total number of objects to 3053.  This set was split into two using a luminosity threshold of $\Lum=10^{27.8}~\mathrm{W~Hz^{-1}}$: the \lp group counts 2593 sources, while the high luminosity one includes 460 of them --- notice that this particular choice of luminosity cut does not affect our results significantly, see~\cite{Pshirkov:2014lqa}: we have also checked this statement directly.  When the $\RMgal$ contribution is subtracted from $\RMobs$\footnote{This step removes the Galactic latitude-dependent contribution from the GMF, which would otherwise contaminate our redshift bins; this also allows us to exploit additional information from the NVSS catalogue.}, one can see that RRMs of sources of lower power have an evolution with redshift consistent with zero.  High power sources are systematically shifted towards higher $\RMin$, but, with the data available today, it was not possible to determine whether this set evolves with $z$.  Table~\ref{tab:bins} summarises this observation for the choices of luminosity cut and redshift binning which were made in~\cite{Pshirkov:2014lqa}.  In this work we employ both the full and the low luminosity sets; we also tested that our results are anyhow robust against the selection of the cut.

\begin{center}
\begin{table}
\resizebox{0.45\textwidth}{!} {
\bgroup
\def\arraystretch{1.5}
\begin{tabular}{c|*{9}{c}}
  $z_b$ & 0.15 & 0.35 & 0.7 & 1.3 & 1.65 & 1.95 & 2.25 & 2.6 & 5 \\ \hline
  $N_s$ & 418 & 418 & 501 & 677 & 291 & 137 & 76 & 50 & 25 \\ \hline
  $\aRMinavg$ & 16.2 & 15.3 & 15.9 & 16.6 & 15.4 & 15.8 & 16.2 & 13.9 & 16.3
\end{tabular}
\egroup}\caption{Upper bin redshift boundaries $z_b$, numbers of sources in the bin $N_s$, and their averages $\aRMinavg$, for the low-power set.}\label{tab:bins}
\end{table}
\end{center}

\paragraph*{Simulations.} When the extragalactic medium is permeated by a MF, this leaves an imprint on the observed RMs, see for example~\cite{Blasi:1999hu,Akahori:2011et}: the $\xRM$ systematically grow with redshift due to their accumulation along the line of sight.  The featureless behaviour which we observed instead is thus incompatible with what is expected if an egMF were present.  By simulating the effects of this hypothesised egMF we can thence constrain its strength.

In order to build a model for $\xRM$ we need to specify the properties of the egMF and the electron density $n_\text{e}$.  Following~\cite{Blasi:1999hu,Akahori:2011et}, the latter is assumed to be well described by the observed Lyman-$\alpha$ (\Lya) forest distribution of neutral hydrogen absorption lines.  In particular, we take the analytical approximation which was given in~\cite{Bi:1996fh,Coles:1991if}, which is a standard log-normal distribution for the electron \emph{overdensity}, $\delta_\text{e}$, with scale parameter
\be
\nonumber
\sigma_\text{e}(z) = 0.08 + \frac{5.37}{(1+z)} - \frac{4.21}{(1+z)^2} + \frac{1.44}{(1+z)^3} \, ,
\ee
and location $\mu_\text{e}(z) = -\sigma_\text{e}^2(z)/2$:
\be
P(\delta_\text{e}) = \frac{1}{\sqrt{2\pi}\sigma_\text{e}(1+\delta_\text{e})} \exp\left\{-\frac{\left[\ln(1+\delta_\text{e})-\mu_\text{e}(z)\right]^2}{2\sigma_\text{e}^2}\right\}\, . \label{ne}
\ee

This distribution is accurate for fluctuations at the Jeans length scale $\lambda_J(z) \simeq 2.3(1+z)^{-3/2}$~Mpc~\cite{Bi:1996fh,Pallottini:2013rja,Choudhury:2000zn,Choudhury:2000gp} (we adopt $H_0 = 71~\mathrm{km/s}/$~Mpc as the Hubble parameter today and $\Omega_M=0.27$ as the total matter density fraction)\footnote{It is possible to adopt different models for $n_\text{e}$, see for instance~\cite{Hui:1996fh}: we found that our results are conservative with respect to this alternative choice.}.  The actual electron density is finally expressed as $n_\text{e}(z) = n_\text{e}(0)(1+\delta_\text{e})(1+z)^3$, with $n_\text{e}(0) = 1.8\times10^{-7}~\mathrm{cm}^{-3}$.

The MF is characterised by the strength $B$ which we wish to constrain, and the power spectrum whose shape is in principle not known.  However, for most cases it can be checked that the effect of the MF on the observed RM is dominated by a single scale, the coherence length $l_c$ which we treat as a free parameter; this scale is not necessarily a physical scale, but a useful practical tool for the analysis: our results are easily rescaled for/adapted to a specific MF model (see~\cite{Pshirkov:2013wka}).  In our simulations we test values for $l_c$ from a tenth of the Jeans length $\lambda_J$ and the Hubble size $1/H_0$.  Since the conductivity of the  Universe is extremely large, and since diffusion of the MF is inefficient at scales much larger than 1~AU, we can safely assume that the MF is frozen into the plasma (neglecting non-linear effects), see for instance~\cite{Durrer:2013pga,Subramanian:2015lua}; thus, for spherical overdensities, it will scale accordingly as $B(z) = \hat{B} \left[n_\text{e}(z)/n_\text{e}(0)\right]^{2/3}$.

Practically, we generate a large number of lines of sight in steps of $\lambda_J$ up to some given redshift, and we collect $\xRM$ from each step.  The electron density is generated at each step sampling the distribution~(\ref{ne}); to simulate the randomness of the MF orientation, we recalculate its amplitude each time the distance travelled equals a multiple of the correlation length, drawing from a uniform $[-1,1]$ distribution~\footnote{Notice that this field is technically not divergenceless; however since the effect of this constraint would be to slightly increase the effective correlation length, and since the latter has a moderate (and certainly subdominant) effect on our results, we can practically ignore this subtlety.}.

\begin{figure}
\begin{center}
  \includegraphics[width=0.48\textwidth]{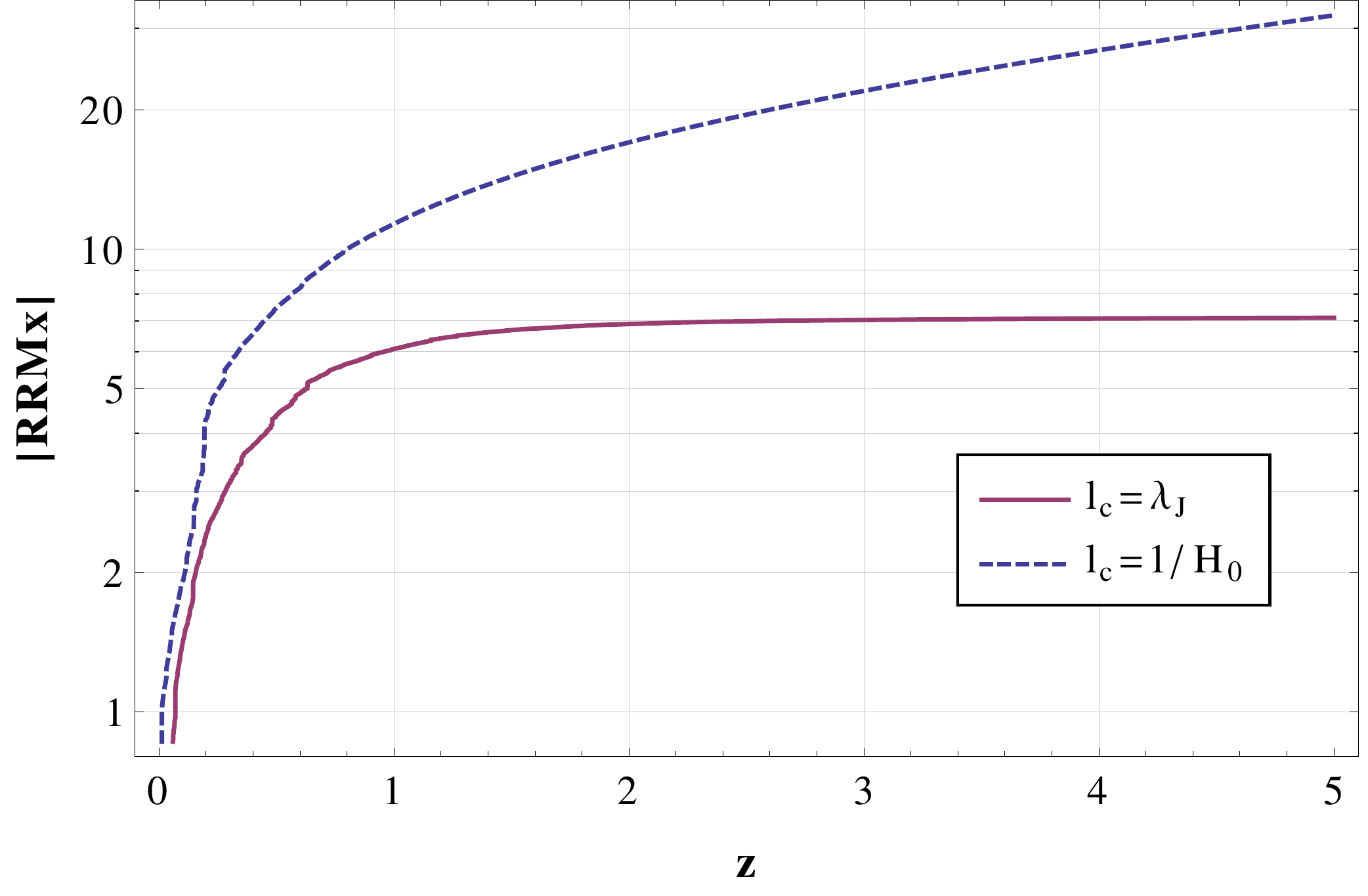}
\end{center}
\caption{Theoretical egMF-induced $|\xRM|$ evolution with redshift for $l_c=\lambda_J$ (red, solid), and $l_c=1/H_0$ (blue, dashed), averaged for $10^5$ lines of sight.}
\label{fig:curve}
\end{figure}

With these prescriptions, we have first obtained the expected theoretical egMF-induced $|\xRM|$ evolution curves with redshift, by averaging $10^5$~lines of sight out to redshift $z=5$: the result is given in Fig.~\ref{fig:curve} for a benchmark MF reference value of $\hat{B} = 1$~nG: the rapid increase of $|\xRM|$ with redshift is evident.  Moreover, there is a clear transition, due to the overall redshift-dependence of $|\xRM|$, from lower to higher redshifts, roughly localised between $z=0.5$ and $z=1$: for $l_c=1/H_0$ the growth of RM is significantly damped at high redshifts, whereas if $l_c=\lambda_J$ the curve becomes essentially flat.  This transition is also observed in the different shapes of the $|\xRM|$ distributions at different redshifts, and is in agreement with the earlier results of~\cite{Blasi:1999hu,Akahori:2011et}.

In our simulations we can generate only the contribution from the egMF.  However, as we mentioned before, when comparing with the data one needs to account also for the turbulent Galactic fields, measurement errors, and intrinsic RMs.  The two first contributions are dominant and do not depend on redshift; the third one does but it is subdominant --- it is precisely this feature, which depend on redshift in a very specific way, that allows us to make this statement, since it is not seen in the data.  We can then exploit the data at low redshifts to obtain information about these redshift-independent contributions, also since at low redshifts they are going to be more important, and then utilise the high-redshift portion of the data to compare with the simulated distributions for the same set of sources, with the egMF included.  For another approach to extract the extragalactic RM piece see~\cite{Oppermann:2014cua}.

We thus split the data in three redshift bands: for simplicity, and because it well matches what we would infer from Fig.~\ref{fig:curve} and more consistently by directly comparing the underlying distributions, we take the low redshift band to correspond to the bins 1 and 2 of Table~\ref{tab:bins}, that is, $z=[0,0.35]$, the high redshift band from $z\geq1.3$ (that is, bins 5 to 9), and a transition band --- which we do not use --- corresponding to bins 3 and 4, or $z=[0.35,1.3]$.  The three sets contain 836 (836 in the \lp set), 1254 (1178), and 936 (579) sources, respectively.  Reducing the size of the intermediate band does not strengthen our constraints but marginally, despite the fact that many sources are found in this redshift range, because the increase in number of sources is counterbalanced by a less pronounced contribution of egMF compared to other sources of $\RMin$.

We finally build the needed \emph{simulated} distribution of $|\RMin|$ at high-$z$ as follows.  First, we randomly pick one $\RMin$ from the low-$z$ set: this serves as our estimation of the $z$-independent contribution which we can not simulate in our model (and for which a specific value we do not want to adopt \emph{a priori}); this is possible because in this low-$z$ bin we have that $|\xRM|\ll|\RMin|$.  In order to obtain the total $|\RMin|$, we then generate a second batch of $\xRM$ values by simulating 100~lines of sight for each of the sources of the high-$z$ set; that is, the $\xRM$ of all high-z sources is simulated 100 times for each source, out to their actual redshift.  This gives a total of 96,300 (57,900)~lines of sight and corresponding $\xRM$.  These $\xRM$ are generated for our benchmark field value of $\hat{B} = 1$~nG, and then rescaled for any other value of $\hat{B}$, simulating the egMF contribution we are seeking.  Randomly picked values from both batches (one each) are then incoherently added (that is, each with its own sign) $10^5$~times to generate the final theoretical $|\RMin|$ distribution, as a function of the MF strength and coherence length, that we can compare with the actual data (strictly speaking this procedure is consistent only for not too small values of the MF strength in order to not introduce ties in the distribution\footnote{Throughout the paper we present our numerical constraints for both the full and \lp data sets; however, we include \lp plots only because they are much cleaner.  If we keep the full set we would introduce some artifacts at low MF strength values (because of artificial ties in the distributions that we feed to the KS test): in the case at hand we have explicitly checked that the relevant p-values are not affected.  We will discuss these (and several other) subtleties at length in a forthcoming publication.}).  Notice that by definition all errors are included in this procedure as we keep the full distribution of $|\RMin|$, not only some of its momenta.

%%%%%%%%%%%%%%%%%%%%%%%%%%%%%%%%%%%%%%%%%%%%%%%%%%%%%%%
\section{Results}\label{sec:res}
%%%%%%%%%%%%%%%%%%%%%%%%%%%%%%%%%%%%%%%%%%%%%%%%%%%%%%%

We compared the two distributions --- the data, and the theoretical predicted sample --- by means of a Kolmogorov-Smirnov (KS) test.  This test allows us to exploit all of the information contained in the distributions, and is more fit to perform this analysis than a simple comparison of means in a given redshift bin, since these tend to fluctuate widely due to the underlying log-normal distribution for $n_\text{e}$, which is indeed the major source of fluctuations.

\begin{figure}
\begin{center}
  \includegraphics[width=0.48\textwidth]{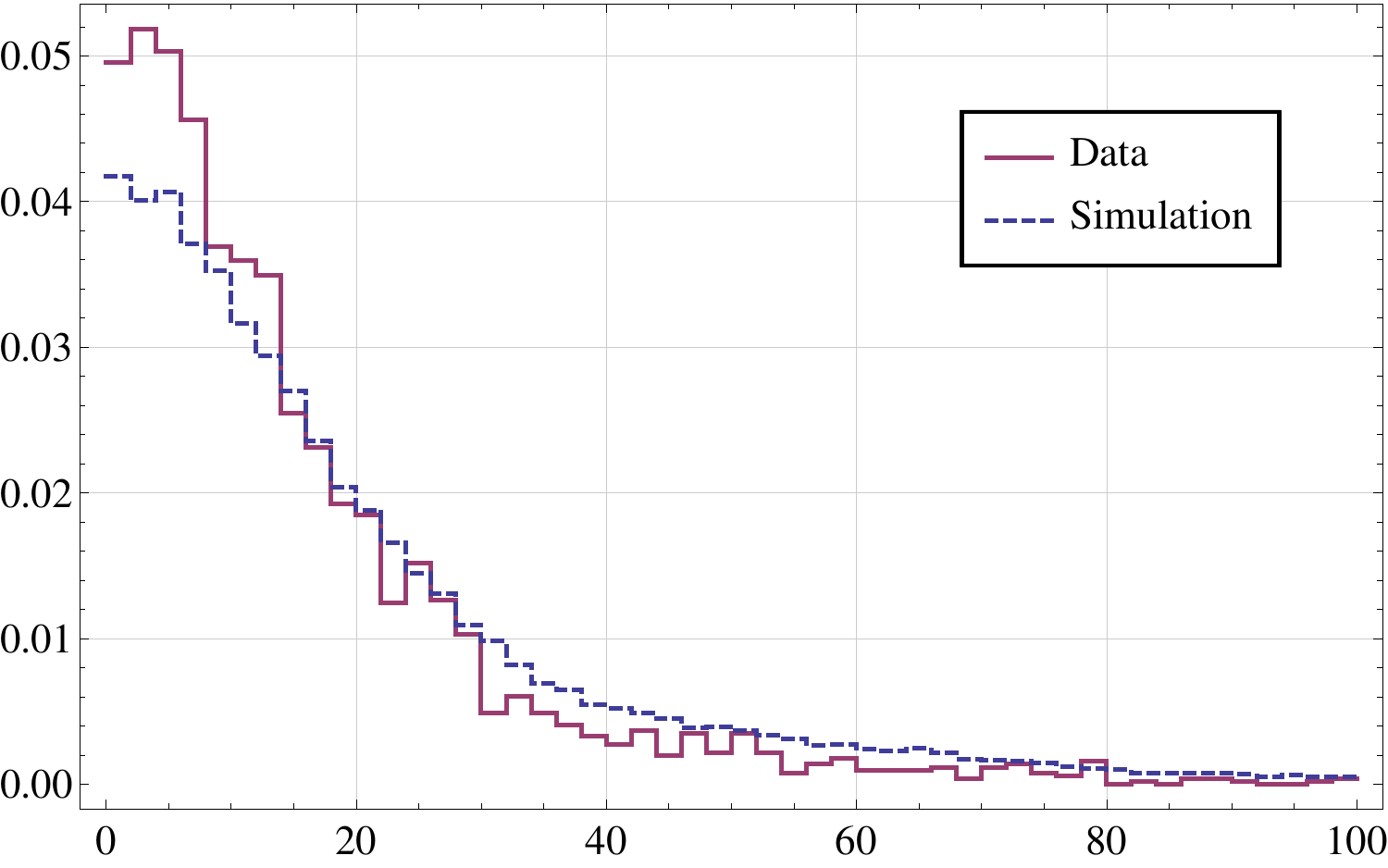}\\
  \vspace{8pt}
  \includegraphics[width=0.48\textwidth]{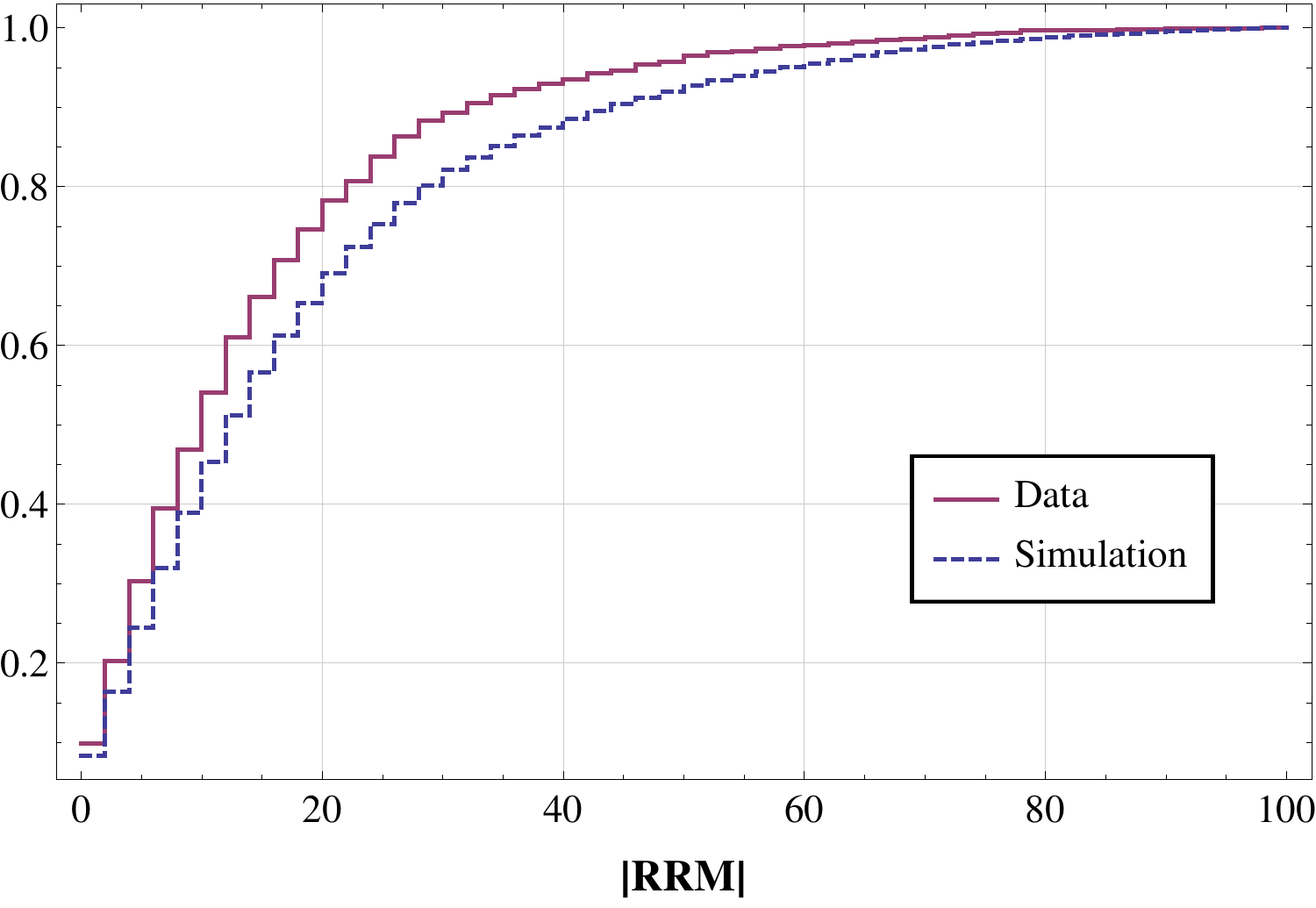}
\end{center}
\caption{PDF and CDF showing $|\RMin|$ for the data (blue, dashed) and a simulated $|\RMin|$ with $\hat{B}=3$~nG and $l_c=\lambda_J$ (red, solid).}
\label{fig:distro}
\end{figure}

In Fig.~\ref{fig:distro} we show an example of the PDFs and CDFs of the two distributions we are comparing: the data (\lp set) and a simulated $|\RMin|$ with $\hat{B}=3$~nG and $l_c=\lambda_J$, where it is clear that such a field value is strongly disfavoured --- the two distributions are statistically incompatible.

\begin{figure}
\begin{center}
  \includegraphics[width=0.48\textwidth]{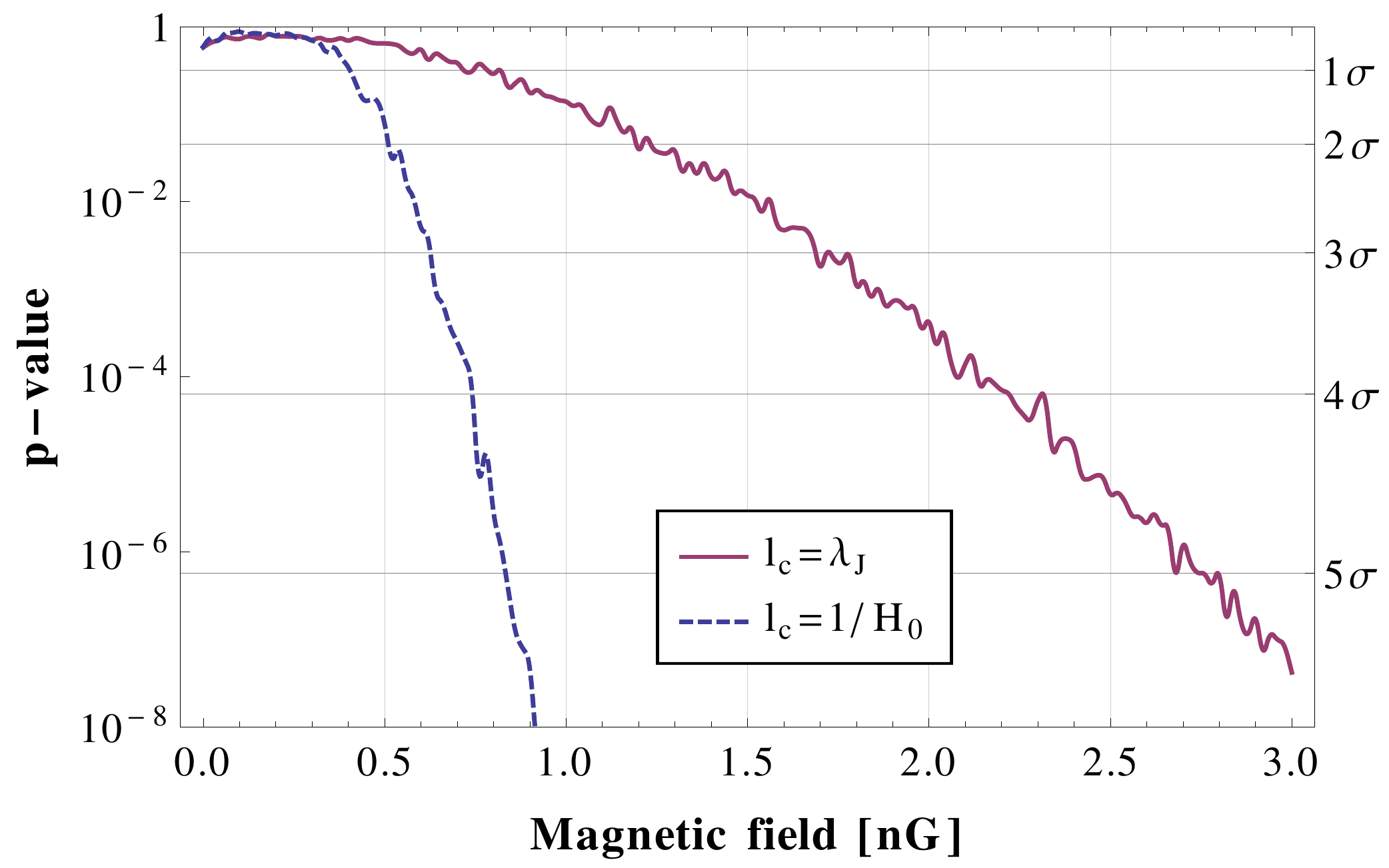}
\end{center}
\caption{KS tests p-values as a function of $\hat{B}$ for $l_c=\lambda_J$ (red, solid), and $l_c=1/H_0$ (blue, dashed).}
\label{fig:ks}
\end{figure}

Fig.~\ref{fig:ks} contains the p-values of the KS tests (\lp set) as a function of $\hat{B}$ for $l_c=\lambda_J$ (red, solid) and $l_c=1/H_0$ (blue, dashed) cases.  For an egMF with $l_c=\lambda_J$ we can read off the values of $\hat{B}$ corresponding to p-values of $2\sigma$ and $3\sigma$ as $\hat{B} = 1.2$~nG, and $\hat{B} = 1.7$~nG, respectively --- these limits read $\hat{B} = 1.7$~nG, and $\hat{B} = 2.2$~nG if we keep the full data set.  For the Universe-wide case the limits are somewhat stronger: $2\sigma$ is already attained at $\hat{B} = 0.50$~nG ($\hat{B} = 0.65$~nG for the full set).  One may worry that in this case however the expected $|\xRM|$ evolution with redshift is not constant, see Fig.~\ref{fig:curve}.  We have checked that the result remains the same if we further split the high redshift band into two because neighbouring high redshifts generate similar $|\xRM|$ distributions.

\begin{figure}
\begin{center}
  \includegraphics[width=0.48\textwidth]{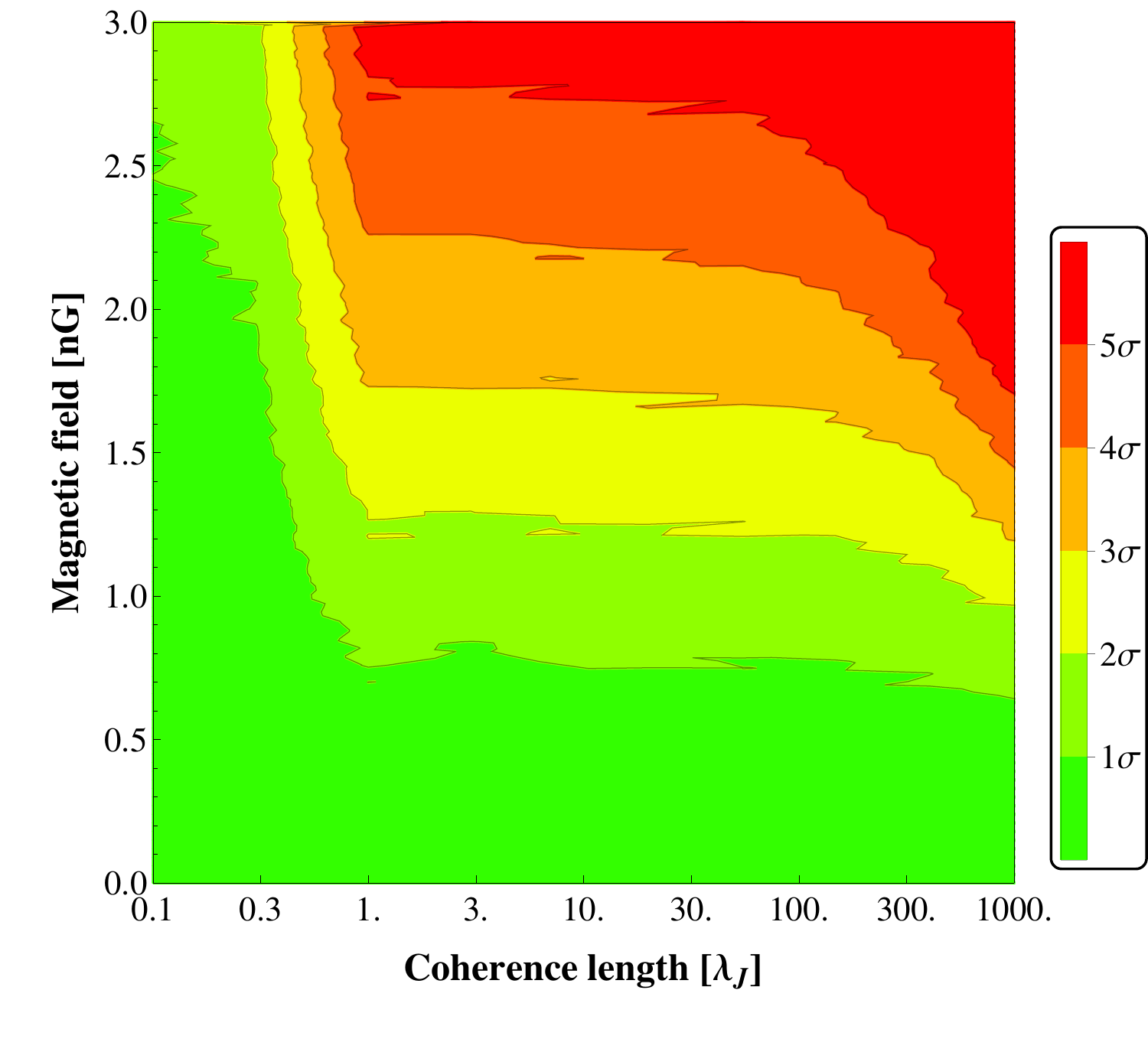}
\end{center}
\caption{Confidence interval contours (bottom panel) obtained from the p-values of the KS tests as a function of $\hat{B}$ (y-axis) and $l_c=\lambda_J$ (x-axis).}
\label{fig:ks2}
\end{figure}

In Fig.~\ref{fig:ks2} we extend the analysis to variable coherence lengths; shown here are confidence interval contours obtained from the p-values of the KS tests as a function of $\hat{B}$ (y-axis) and $l_c=\lambda_J$ (x-axis).  We see here how the limits can be automatically rescaled for $l_c<\lambda_J$, since $n_e$ fluctuations are dominated by the $\lambda_J$ wavelength: the bound on the field strength becomes approximately $(\lambda_J/l_c)^{1/2}>1$ times weaker (that is, the allowed egMF is $(\lambda_J/l_c)^{1/2}>1$ times stronger).

%%%%%%%%%%%%%%%%%%%%%%%%%%%%%%%%%%%%%%%%%%%%%%%%%%%%%%%
\section{Conclusion}\label{sec:end}
%%%%%%%%%%%%%%%%%%%%%%%%%%%%%%%%%%%%%%%%%%%%%%%%%%%%%%%

Extragalactic MF with coherence lengths of about 1~Mpc can not be stronger than 1.7~nG at the $2\sigma$ level (1.2~nG for the low-luminosity data), whereas a Universe-wide egMF is bound to be weaker than about 0.65~nG (0.50~nG).  These limits are obtained using RM data from extragalactic sources, are valid independently of the origin of these egMF, and for a very large class of egMF models.  Moreover, these limits are a fivefold improvement over those previously available in the literature, and are now more than competitive with microwave background ones (which however apply only to primordial fields).

This improvement stems from the observation that RMs of distant objects do not evolve with redshift.  This, combined with a much larger and better set of data, the RM compilation of the NVSS catalogue, enabled a much more robust statistical approach to constraining the egMF.

A straightforward application of these limits is to UHECRs propagation: if an egMF with $\hat{B} \simeq 1$~nG and $l_c \simeq \lambda_J$ existed, it would alter the way UHECRs propagate by deflecting them quite significantly.  The median deflection for a proton primary of even the highest energy, $10^{20}$~eV, would be around 9~deg when propagating from a distance of 200~Mpc.  This is a result of a simulation with the same \Lya~distribution for $n_\text{e}$ and $\hat{B}$ rather than the simplified homogeneous cell model usually implemented.

Our approach can be directly ported to analyse the data coming from the next generation of radio telescopes, in particular the Square Kilometer Array (SKA)~\cite{Carilli:2004nx}. The main limitation in the currently available data is the low number of sources, particularly those with precise redshift and spectral index information.  The new data will significantly overcome this limitation, and allow a much more robust determination, and a marked strengthening, of our limits; we discuss the sensitivity of our new technique in the forthcoming SKA era in a future publication.

%%%%%%%%%%%%%%%%%%%%%%%%%%%%%%%%%%%%%%%%%%%%%%%%%%%%%%%
\paragraph*{Acknowledgements.} We would like to thank Andrei Gruzinov for useful comments.  FU and PT are supported by IISN project No.~4.4502.13 and Belgian Science Policy under IAP VII/37. PT is supported in part by the RFBR grant 13-02-12175-ofi-m. MP are supported by  the Russian Science Foundation grant 14-12-01340.  MP acknowledges the fellowship of the Dynasty Foundation, and want to thank the Service de Physique Th\'eorique in Brussels, where this work was completed.

\bibliography{eXg}

\end{document}